\documentclass[aps,prl,twocolumn]{revtex4}
\usepackage{graphicx,amssymb}
\usepackage{epsf}  

\begin{document}
\title{Evidence of collective charge transport in the Ohmic regime of $o$-TaS$_3$ in the charge density wave state by a photoconduction study} 

\author{S.V. Zaitsev-Zotov and V.E. Minakova}
\affiliation{Institute of Radioengineering and Electronics of RAS, 125009
Mokhovaya 11, Moscow, Russia}

\begin{abstract} Using photoconduction study we demonstrate that the low-temperature Ohmic conduction of o-TaS$_3$ is not provided by band motion or hopping of  single-particle excitations -- electrons and holes excited over the Peierls gap. Instead, the low-temperature Ohmic conduction is mostly provided by collective excitations having the activation energy much less than the Peierls gap value and shunting the contribution of electrons and holes responsible for photoconduction.
\end{abstract}

\maketitle

A charge-density wave (CDW) developing  below the Peierls transition temperature in a wide class of solids with a chain-like structure and nearly one-dimensional electron spectrum determines many unusual properties of these materials \cite{cdwreview}. Sliding of  the CDW along a crystal provides a collective contribution into conduction. Interaction of the CDW with impurities leads to pinning. As a result, the CDW contribution appears in the electric field, $E$,  exceeding a threshold value $E_T$. At $E\lessapprox E_T$ the collective contribution of the CDW into charge transport occurs through thermally activated creeping. Conduction of the Peierls conductors at $E\ll E_T$ is determined by single-electron excitations (electrons and holes) thermally excited over the Peierls gap. Such a point of view can be found in many papers describing properties of  the Peierls conductors. It will be shown below, that at least in o-TaS$_3$ at temperatures below 100~K  the dominant contribution into the linear conduction is not provided by single-particle mechanisms.

Orthorhombic o-TaS$_3$ is a typical quasi-one-dimensional conductor. A temperature dependence of the linear conductivity in this material below the Peierls transition temperature $T_P=220$~K  obeys the activation law with an activation energy $\Delta= 850$~K. At $T\lessapprox T_P/2$ the linear conductivity begins to deviate from the activation law, and the activation energy is getting smaller \cite{sambongi}. Similar deviation is observed in many CDW materials, such as blue bronzes, (TaSe$_4)_2$I {\it etc.} \cite{cdwreview}, and in spin-density wave (SDW) conductors \cite{sdwreview}. Since the first publication on transport properties of o-TaS$_3$ \cite{sambongi} this deviation from the activation law is often attributed to contribution of collective excitations (phase or amplitude solitons), whereas no experimental evidence has been provided so far that it is not due to single-particle excitations moving in a spatially-nonuniform potential relief caused by pinned CDW or hopping via impurity levels. Temperature dependence of transverse conduction \cite{sambongi} is often considered as an argument in favor of this point of view: it follows the activation law even at $T<T_P/2$. Recent measurements \cite{transpok} demonstrate, however, that the transverse conduction, including the earlier results, is better described by the variable-range hopping mechanisms, so this argument is not a straightforward one. Numerous experimental attempts to detect solitons by optical methods \cite{brill} do not provide a convincing evidence of their existence in o-TaS$_3$.

Photoconduction (illumination-induced change of the {\em linear} conduction) was observed recently in thin crystals of o-TaS$_3$ \cite{zzminajetpl}. This phenomenon gives us a tool to investigate the nature of low-temperature linear conduction in o-TaS$_3$. By using this tool we demonstrate here that at low temperatures the photoconduction and conduction of o-TaS$_3$ are provided by different physical mechanisms. Namely, photoconduction is mostly provided by nonequilibrium electrons and holes optically excited over the Peierls gap, whereas the low-temperature conduction is  not of a single-particle origin and is presumably related to collective excitations of the CDW, i.e. to solitons. 

Photoconduction was studied in relatively thin crystals of o-TaS$_3$ with a unit length resistance value $R_{300}/L=(0.02{\rm -} 1.3){\rm ~}k\Omega /\mu$m corresponding to the cross-section areas $2\times 10^{-3}-2 \times 10^{-2}$~$\mu$m$^2$. The physical properties of such crystals are still very close to those of bulk ones \cite{zzreview,zzmicroengelectr}.  At the same time, as their transverse sizes are comparable or even smaller the light-penetration depth ($0.1{\rm -}1$~$\mu$m \cite{itkis}), so photoexcitation of electrons and holes over the Peierls gap is almost spatially uniform across the sample cross-section. Current terminals to the crystals were made by indium cold soldering. All measurements were performed in two-terminal configuration in the voltage-controlled regime at $E\leq 0.2 E_T$. IR LED with 1.16~eV photon energy above the optical value of the Peierls gap of o-TaS$_3$ ($250$~meV \cite{ogap}) was used for illumination. 

Earlier measurements of intensity dependent photoconduction $\delta G$ done with the steady light illumination \cite{zzminajetpl} gave a very complex picture of the phenomenon.  $\delta G(W,T)$ dependence,  where  $\delta G(W,T)\equiv G(W,T)-G(0,T)$, and $W$ is the light intensity, was found to be very nonlinear function of $W$ with temperature dependent parameters. Photoconduction kinetics study results reported here provide more simple picture and give us a clue for explanation of many features of photoconduction and transport properties of o-TaS$_3$. 

\begin{figure} 
\epsfxsize=9cm
\leavevmode{\epsfbox{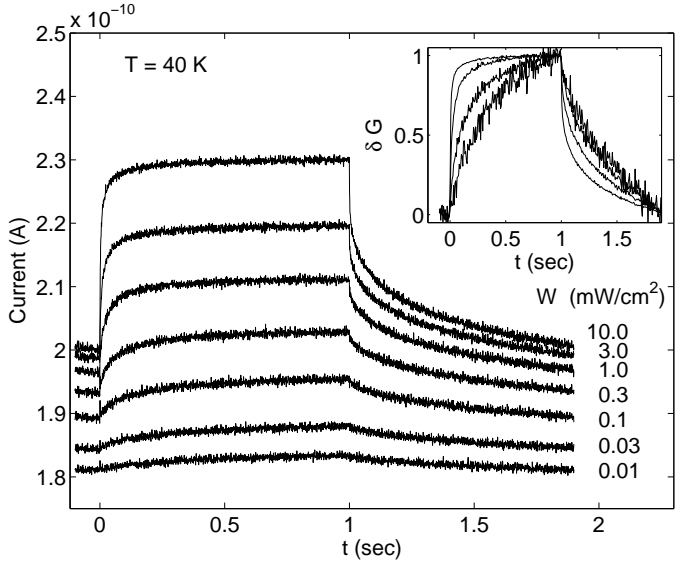}}
\caption[]{ Photoresponse of o-TaS$_3$ sample to a meander light pulse sequence 
(light is on at $t=0$ and off at $t=1$~s)  at different light intensities. 
$T=40$~K, $V=200$~mV. Sample parameters: $L=0.4$~mm, $R_{300K}=7$~k$\Omega$. Inset shows the shape of the pulse 
$\delta I(t)/\delta I(1{\rm s})=[I(t)-I(0)]/[I(1{\rm s})-I(0)]$ at $W=10$, 1, 0.1 and 0.01 mW/cm$^2$. }
\label{relht}
\end{figure}

Fig.~\ref{relht} shows typical time evolution curves of photoconduction of o-TaS$_3$. Photoresponse at $T\gtrsim 40$~K is relatively slow with intensity-dependent relaxation time of millisecond region. No fast contribution with expected picosecond relaxation time \cite{comment} is observed.  Inset in Fig.~\ref{relht} shows the same curves normalized to their amplitudes. The shape of the curves depends on the light intensity and relaxation is getting faster for more intensive light. The photocurrent is a nonlinear function of the light intensity: {\it e.g.} growth of $W$ from 0.1 to 0.3 mW/cm$^2$ results in approximately the same increase of $I$ as growth of $W$ from 0.3 to 1 mW/cm$^2$ (Fig.~\ref{relht}).

\begin{figure} 
\epsfxsize=9cm
\leavevmode{\epsfbox{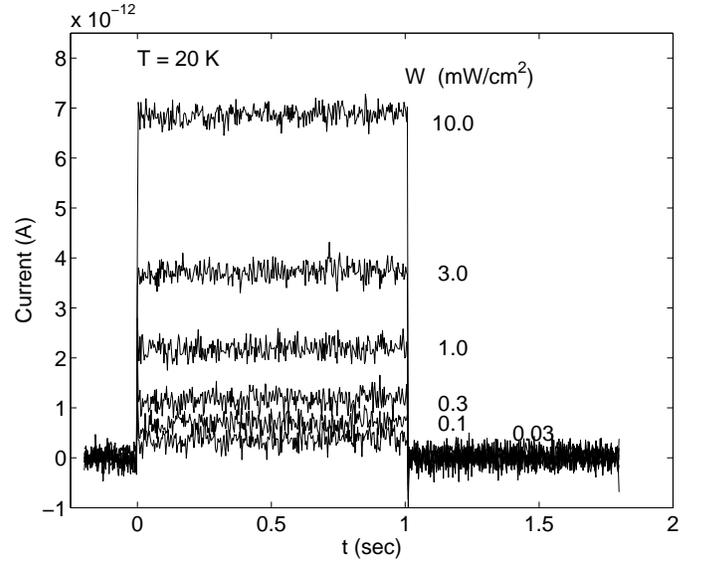}}
\caption[]{Photoresponse of o-TaS$_3$ sample to a meander light pulse sequence 
(light is on at $t=0$ and off at $t=1$~s)  at different light intensities.   Sample parameters: $L=0.4$~mm, $R_{300K}=7$~k$\Omega$. $T=20$~K, $V=2$~V.}
\label{rellt}
\end{figure}

The intensity range where slow response is observed is shifted to smaller light intensities upon cooling, and the response at a fixed light intensity is getting faster.  Finally, at sufficiently low temperatures the photoconduction relaxation time is much faster than the time constant of our amplifier (0.3 ms) and the shape of the response does not look intensity dependent (Fig.~\ref{rellt}). Detailed study of temperature evolution of photoconduction kinetics proves that the fast low-temperature response can be considered as the limiting case of very high light intensity.

\begin{figure} 
\epsfxsize=9cm
\leavevmode{\epsfbox{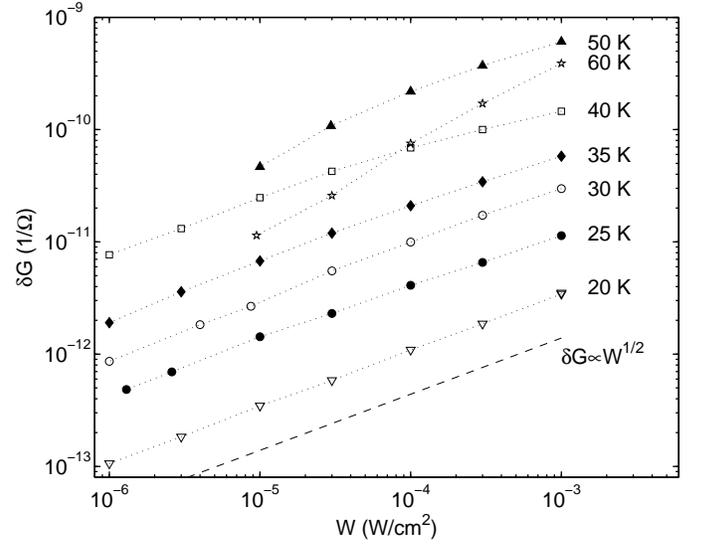}}
\caption[]{An amplitude of photoresponse of o-TaS$_3$ sample to a meander light pulse sequence (light pulse duration 1 s, a period 2 s) {\it vs.} light intensity at different temperatures. The dashed line shows the slope of $\delta I\propto W^{1/2}$ expected for the quadratic recombination mechanism. Sample parameters: $L=0.4$~mm, $R_{300K}=7$~k$\Omega$.}
\label{isdall}
\end{figure}

The amplitude of photoconduction measured by the light modulation method is found to be a nonlinear function of the light intensity (Fig.~\ref{isdall}). At $T\lesssim 50$~K it can be approximated as $\delta G\propto \sqrt(W)$, in contrast to $\delta G\propto W^\alpha$ with temperature dependent $\alpha$ ($0.2 \lessapprox \alpha \lessapprox 1$) for steady light illumination \cite{zzminajetpl}. This law corresponds to so-called quadratic recombination which is intrinsic for semiconductors in the high-intensity limit $\delta G \gg G$ \cite{ryvkin,moss}. In semiconductors the low-intensity limit ($\delta G \ll G$) is characterized by the linear relation $\delta G \propto W$, and the crossover between the linear and quadratic recombination occurs at $\delta G\sim G$. Therefore, to determine the carriers responsible for photoconduction in o-TaS$_3$, it is worth to search for the linear recombination region at much lower intensities. 

\begin{figure}
\epsfxsize=9cm
\leavevmode \centering{\epsfbox{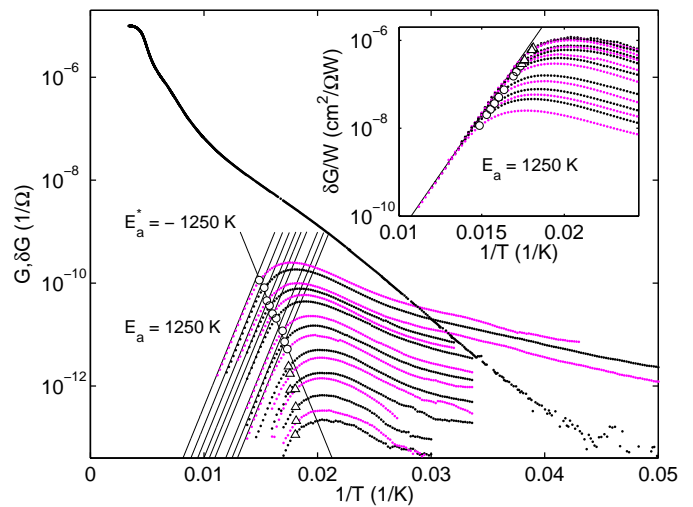}}
\caption[]{Temperature variation of the linear conductance of o-TaS$_3$ in the dark (upper curve). Set of curves shows temperature dependencies of AC photoconduction measured at light intensities (from top to bottom) 10, 4, 1.8, 1, 0.5, 0.28, 0.077, 0.038, 0.02, 0.0082, 0.005, 0.0024, 0.0014, 0.00062, 0.00031, and 0.00019~mW/cm$^2$.  The lines corresponds to the activation law with the activation energies $E_a=1250$~K and $E^*_a=-1250$~K.
Inset shows the same set of $\delta G$ curves normalized by the light intensities. The solid line results from collapse of  the activation dependences with $E_a=1250$~K shown by lines in Fig.~\protect\ref{rtid}.
Sample parameters: $L=0.7$~mm, $R_{300K}=95$~k$\Omega$. $f=4.5$~Hz. See text for details.}
\label{rtid}
\end{figure}

Fig.~\ref{rtid} shows temperature dependence of the linear conductance of o-TaS$_3$ together with AC photoconductance $\delta G$ measured by the standard light modulation technique in the temperature range  $20 \leq T\leq 300$~K. $G(T)$ curve is usual for this material. It consists of the high-temperature metallic part, the Peierls transition at 210~K (shifted due to the finite-size effect \cite{zzreview}), activation region with the activation energy 850~K corresponding to the half of the high-temperature value of the Peierls gap $2\Delta$, and the low-temperature region with the activation energy about  400~K.  

To find the level of linear conduction for current carriers responsible for photoconduction we have to analyze $\delta G(T)$ curves more accurately. Inset in Fig.~\ref{rtid} shows the same set of $\delta G(T)$ curves normalized to the light intensity values. At $T\geq 60$~K all  
$\delta G(T)/W$ curves collapse into a single curve which can be fitted by the activation law $\delta G(T)/W\propto \exp (E_a/kT)$ with the activation energy $E_a=(1250\pm 50)$~K. The temperature of deviation from this activation law  decreases with decrease of  $W$. Similarly, normalization by $\sqrt{W}$  collapses $\delta G(T)/\sqrt{W}$ curves in the low-temperature region below 40 К, in agreement with Fig.~\ref{isdall}. Thus, the maxima of $\delta G(T,W)$ dependences separate linear and nonlinear recombination regimes. 

The concentration of electrons in semiconductors with quadratic recombination obeys the  equation \cite{ryvkin,moss}
\begin{equation}
\frac{dn}{dt}= \frac{k\beta W}{\hbar\omega}-R(np-n_0p_0),
\label{dndt}
\end{equation}
 where $k$ is the quantum efficiency of photogeneration, $\beta$ is the absorption coefficient,  $n_0$ and $p_0$, $n$ and $p$ are respectively the electron and hole concentrations in and out of equilibrium, $R$  is a recombination coefficient and $W$ is a light intensity. The  equation for hole concentration is similar.  Let us assume for simplicity that $n_0=p_0$, $\delta n=\delta p$, where $\delta n=n-n_0$, $\delta p=p-p_0$. In the stationary state $dn/dt=0$. We are looking for $\delta n/W$ values for two cases: i) when $\delta n = n_0$, and ii) when $W\rightarrow 0$.  From equation \ref{dndt} one gets: i) $\delta n/W=k\beta /3\hbar \omega n_0R$ and ii) $\delta n/W=k\beta /2\hbar \omega n_0R$. So the ratio $\delta n/W$ at $\delta n=n_0$ is 1.5 times smaller than its value at  $W\rightarrow 0$.
 
In nonstationary case the signal amplitude depends also on the modulation frequency $f$. The conduction does not reach its stationary value when the light is on and does not relax to its equilibrium value when the light is off (see, e.g. Fig.~\ref{relht} at $W\leq 0.1$ mW/cm$^2$). The measured AC conduction modulation is therefore only a fraction of  photoconduction of the stationary case. For data presented in  Fig.~\ref{rtid} ($f=4.5$~Hz) the crossover to nonstationary case takes place around 60~K, so $\tau (60{\rm~K}) \sim 1/2f\simeq 0.1$~s. At $T< 60$~K  $f\gtrsim 1/2\tau$, i.e. photoconduction measured at small light intensities is nonstationary. 

Open circles and triangles in the inset in Fig.~\ref{rtid} mark  $\delta G/W$ level which is 1.5 times smaller than the activation one for stationary (circles) and nonstationary (triangles) photoconduction.  As it was argued above, in the stationary case the photoconduction $\delta G$ at these points is equal to the linear conduction $G_0$ of photoconduction-providing current carriers. Corresponding values of $G_0$ are also shown in Fig.~\ref{rtid} by the same markers. The high-temperature part of the curve (circles, $T\geq 60$~K) follows the activation dependence $G_0(T)\propto \exp (E^*_a/kT)$ with the activation energy $E^*_a\approx -E_a$, and is much steeper  in the nonstationary case (triangles).
 
 The relaxation time can be found from the amplitude of photoconduction in the stationary state (we assume for simplicity that $1/\beta$ is much smaller than our sample thickness \cite{itkis})
 \begin{equation}
 \tau=\frac{\delta n \hbar \omega}{k \beta W}.
\label{tau}
 \end{equation}
As  $k$ and $\beta$ are not expected to be temperature dependent, so in the linear relaxation regime $\delta G(T)/W\propto \tau(T)$ (inset in Fig.~\ref{rtid})  \cite{comment_tau}. The characteristic time, $\tau_\alpha$, of the main dielectric relaxation process ($\alpha$-process) fitted in Ref.~\cite{bel} by the Vogel-Fulche law, at $60\leq T\leq 100$~K has the same temperature dependence as  $\tau(T)$  observed here.  $\tau_\alpha(T)$  was attributed to the temperature variation of quasiparticle concentration \cite{bel}. Similarly,  $\tau(T)$ also follows quasiparticle concentration (see below). It is very likely, therefore, that these relaxations represent two sides of the same process of electron redistribution.

Recent interlayer tunneling experiments \cite{lat} revealed step-like temperature dependence of the tunneling gap  in o-TaS$_3$ indicating appearance of a new low-temperature phase  below 100~K,  presumably corresponding to the commensurate CDW \cite{lat}. This phase has a bigger value of the gap,  $2\Delta^*\approx 2300$~K, than the high-temperature phase ($2\Delta=1700$~K),  $\Delta^*$ being very close to both $E_a$ and $-E^*_a$. In its turn, the interband radiative recombination lifetime $\tau_r$ (the process responsible for quadratic recombination) of semiconductors depends on the Fermi level position. In pure semiconductors  $\tau_r\propto \exp (E_G/2T)$, where $E_G$ is a semiconductor energy gap \cite{ryvkin,moss}. In doped semiconductors $\tau_r$ is much smaller and has much weaker temperature dependence \cite{ryvkin,moss}. In o-TaS$_3$ ($E_g=2 \Delta^*$) $E_a\approx E_g/2$ . Thus, photoconduction in o-TaS$_3$ is provided by electrons and holes excited over the Peierls gap, and the Fermi energy is in the middle of the Peierls gap. 

The origin of the low-temperature linear conduction with the activation energy much smaller the Peierls gap value is an adjacent question of interest. In principle, an additional conduction channel provided by hopping may shunt conduction of electrons and holes thermally excited over the Peierls gap. Hopping by impurity levels requires the Fermi level at impurity level \cite{hopping}, i.e. about 400~K far from the Peierls gap edge, in contradiction with our conclusion about its middle-gap position. The variable-range-hopping mechanism implies a nonzero density of states near the Fermi level (non-zero for Mott-type conduction, and with a dip at the Fermi level for Efros-Shklovskii conduction  \cite{hopping}). No middle-gap features are observed in the tunneling  study \cite{lat}. Hence,  the low-temperature linear conduction of o-TaS$_3$ is not provided by electrons and holes and is, therefore, of collective origin. This collective contribution dominates in the low-temperature linear conductivity and shunts the contribution of electrons and holes excited over the Peierls gap.

The possibility to distinguish between contributions of single-particle and collective excitations of the CDW (solitons) to the linear conduction is the most interesting result of the present study. We would like to note that   photoexcitation of nonequilibrium electrons and holes may also change the soliton concentration and provide thereby a collective contribution into photoconduction which can be considered as a ``collective photoconduction'' \cite{collective}. Interrelationship between the collective and single-particle photoexcitations is still an open question.

As it was mentioned above, the low-temperature deviation of $G(T)$ from the activation law is a feature of many CDW and SDW materials. The method described here may help to clarify of the origin of this deviation in these materials, including, for example, the blue bronze K$_{0.3}$MoO$_3$ where photoinduced changes of {\it nonlinear} conduction and CDW creep rate are  observed \cite{collective,ogawa}. 

In conclusion, it is shown that photoconduction in quasi-one-dimensional conductors can be used as a probe of single-particle conduction. Our results demonstrate the presence of low-temperature collective conduction shunting single-particle conduction in o-TaS$_3$.  

\acknowledgments
We are grateful to S.~N.~Artemenko and V.~Ya.~Pok\-rovskii for useful discussions 
and R.~E.~Thorne for providing high-quality crystals. The work was supported RFBR.
These researches were performed in the frame of the CNRS-RAS-RFBR Associated 
European Laboratory ``Physical properties of coherent electronic states in 
condensed matter'' between CRTBT and IRE.



\begin{thebibliography}{99}
\bibitem{comment_tau}Taking $\delta n\sim n_{300} \delta G/G_{300}$,  where $n_{300}=2\times 10^{21}$~cm$^{-3}$ and $G_{300}=1.05\times 10^{-5}$~1/$\Omega$ are respectively electron concentration and conductance at $T=300$~K,  $\delta G({\rm 60~K})/W= 3 \times 10^{-7}$~cm$^2$/$\Omega$W,  $1/\beta=1$~$\mu$m, and $k=1$, one gets a rough estimate $\tau (60{\rm ~K}) \sim 10^{-3}$ s, much closer to observed value 0.1 s, than the theoretical prediction \cite{brazovskii} and the blue bronze data  \cite{bbrelax}.
\end{thebibliography}
\end{document}